\documentstyle[11pt,newpasp,twoside]{article}
\def\edcomment#1{\iffalse\marginpar{\raggedright\sl#1\/}\else\relax\fi}
\reversemarginpar

\newcommand\arcdeg{\mbox{$^\circ$}}%
\newcommand{\hb}{H\,$\beta$}
\newcommand{\mgii}{Mg\,{\sc ii}}
\newcommand{\oiii}{[O\,{\sc iii}]}
\newcommand{\niv}{N\,{\sc iv}]}
\newcommand{\civ}{C\,{\sc iv}}
\newcommand{\ciii}{C\,{\sc iii}]}
\newcommand{\siiv}{Si\,{\sc iv}}

\newcommand{\heii}{He\,{\sc ii}}
\newcommand{\feii}{Fe\,{\sc ii}}
\newcommand{\caii}{Ca\,{\sc ii}}
\newcommand{\lam}{$\lambda$}
\newcommand{\mbh}{$M_{\rm BH}$}

\newcommand{\lsim}{\stackrel{\scriptscriptstyle <}{\scriptstyle {}_\sim}}

\newcommand{\et}{\mbox{et~al.\ }}
\newcommand{\eg}{\mbox{e.g.,}\ }

\begin{document}
\title{Black-Hole Mass Measurements}
\author{M. Vestergaard}
\affil{Department of Astronomy, The Ohio State University, 140\,West\,18th Avenue, Columbus, OH\,43210} 

\begin{abstract}
The applicability and apparent uncertainties of the techniques currently available 
for measuring or estimating black-hole masses in AGNs are briefly summarized.
\end{abstract}

\section{Introduction}
Knowledge of the mass of the central black hole in active galaxies
is important for many studies of early structure formation and of
the central engine and its evolution. The hope is that we will learn
how the formation and growth of the black hole affects the formation,
the evolution, and the characteristics of the galaxy by which it is hosted.
The SDSS promises to provide valuable insight.

Black-hole mass determination methods fall in two distinct categories:
primary and secondary methods. Their differences are important to recognize.
With primary methods, the black-hole mass (\mbh) is directly {\sl measured} from gas
and stars whose dynamics are dictated by the black hole. These methods are
often very challenging or time consuming. In contrast, secondary methods
are often more easily applied to large data sets, but 
only provide {\sl estimates} of the mass by adopting approximations to the
primary methods or by measuring parameters with which \mbh{} 
is known to correlate. 

Due to space limitations, all the work done on this subject cannot be cited; 
my apologies in advance.  Emphasis is placed on a few key papers
providing calibration of the methods and assessments 
of the uncertainties.

\section{Primary Mass Determination Methods} 

\begin{table}[t]
\caption{Primary Mass Determination Methods}
\begin{center}
\begin{tabular}{lcccc}
\\[-11pt] \tableline \\[-3pt]
&\multicolumn{2}{c}{\bf Low-\boldmath{$z$}} & {\bf High-\boldmath{$z$}} & {\bf Best} \\
& {\bf Low-\boldmath{$L$}} & {\bf High-\boldmath{$L$}} & {\bf High-\boldmath{$L$}} & {\bf Accuracy} \\[1pt]
& LINERs, & QSOs, Sy\,1s & QSOs & (dex) \\[1pt]
& Sy\,2s  & BL Lacs & & \\[3pt]
\tableline \\[-3pt]
Stellar \& Gas & \\
\hspace{0.2cm} kinematics     & ($\surd$) & $\div$ & $\div$ & 0.15\,$-$\,0.3 \\[1pt]
Megamasers    & Type\,2   & $\div$ & $\div$ & $\leq -1.0$ \\[1pt]
Reverberation  &         &         &         &            \\[1pt]
\hspace{0.2cm} 
  Mapping        & Type\,1 & Type\,1 & $\surd$ & 0.15\,$-$\,0.3 \\[1pt]
\tableline \tableline
\end{tabular}
\end{center}
\vspace{-0.5cm}
\end{table}

\paragraph{Stellar and Gas Kinematics.\ } 

To determine the virial black-hole mass of nearby quiescent galaxies 
requires high spatial resolution spectroscopy of the nuclear regions
in order to measure the velocity dispersion of the stars (or gas) 
within the sphere of influence of the central black hole. This method 
can also be used for weakly active galactic nuclei (AGNs), such as 
low-luminosity AGNs, LINERs, and Seyfert~2s (Table~1). For more luminous 
AGNs (Type~1 sources) the strong nuclear emission washes out the stellar
features in the spectrum. The scatter
in the \mbh{}$- \sigma_{\ast}^{\rm bulge}$ relationship, established 
for nearby quiescent galaxies (\eg Tremaine \et 2002), suggests that 
black-hole masses determined from stellar and gas kinematics are accurate 
to within a factor of about 2.

\paragraph{Megamasers.\ } 

Spectroscopy of water-vapor maser emission from circumnuclear
disks in nearby AGNs reveals the %(Keplerian) 
kinematics of the disk and the location of the maser.  
Megamasers can yield highly accurate \mbh{} measurements, especially in
highly inclined sources: for NGC\,4258, which is inclined 
83\arcdeg$\pm$4\arcdeg, %83$\pm$4 degrees, 
the mass is determined to within a few
per cent (Miyoshi \et 1995). However, this method is only useful for 
selected `edge-on' sources and so far relatively few 
objects are known to have megamasers (\eg Greenhill \et 2003).

\paragraph{Reverberation Mapping.\ } 

The reverberation-mapping technique is the best and most robust mass 
measurement method to apply to AGNs and quasars: utilizing the 
variability properties of the source the technique does not require
high spatial resolution and is not affected by the nuclear glare.
Peterson \& Onken (2004, this volume) outline the basic principle of the 
reverberation mapping technique and discuss the importance of the
zero-point calibration of reverberation masses and of the unknown,
order unity, scale factor $f$. 

For a given AGN the {\sl rms} velocity width and time lag for different
broad emission lines correlate such that higher ionization lines have
larger widths and smaller lags (consistent with the same central mass for 
each object). This virial relationship for multiple 
lines is seen for all four AGNs for which this is testable (NGC\,7469, 
NGC\,3783, NGC\,5548, 3C\,390.3)
and for the well-measured emission lines of \siiv{} \lam
1400, \civ{} \lam 1549, \heii{} \lam 1640, \ciii{} \lam 1909, \hb{} \lam 
4861, and \heii{} \lam 4686
(Peterson \& Wandel 1999, 2000; Onken \& Peterson 2002). 

Notably, nearby AGNs with measurements of both reverberation mass and 
bulge velocity dispersion fall along the \mbh{} $- \sigma_{\ast}$ 
relationship established by quiescent galaxies and with a similar scatter 
(Ferrarese \et 2001). This suggests \mbox{that reverberation mapping masses are 
accurate to within a factor 2 to 3.}

Reverberation mapping is fully applicable to distant quasars (Table~1),
but is less practical as it is extremely time and resource consuming: 
luminous AGNs %quasars
vary with smaller amplitudes and on longer time scales than %the 
lower-luminosity, nearby AGNs. The %long 
time scales are further
increased by time dilation owing to the cosmological distances of quasars.
Even for the 17 nearby quasars monitored so far, %some 
ten years of variability data are necessary to obtain reasonable 
%reverberation 
results (Kaspi \et 2000). 
Secondary methods are more practical for distant quasars. 

\begin{table}[t] 
\caption{Secondary Mass Determination Methods}
\begin{center}
\begin{tabular}{lccccl} 
\\[-28pt] \tableline \\[-3pt]
&\multicolumn{2}{c}{\bf Low-\boldmath{$z$}} & {\bf High-\boldmath{$z$}} & {\bf Best} & {\bf Future}\\
& {\bf Low-\boldmath{$L$}} & {\bf High-\boldmath{$L$}} & {\bf High-\boldmath{$L$}} & {\bf Accu-} & {\bf Work} \\[1pt]
& LINERs, & QSOs,\,Sy\,1s & QSOs & {\bf racy} & \\[1pt]
& Sy\,2s  & BL Lacs & & (dex) & \\[3pt]
\tableline \\[-3pt]
Scaling Relations & $\surd$ & $\surd$ & $\surd$ & 0.4\,$-$\,0.5 & note $a$ \\
Via \mbh{}\,$-$\,$\sigma_{\ast}$:  & \\
\hspace{0.2cm}
 $-$ $\sigma_{\ast}$ & $\surd$ & $\surd$ & $\div$ & 0.3 & note $b$ \\[1pt] 
\hspace{0.2cm}
 $-$ \oiii{} FWHM & $\surd$ & $\surd$ & $\surd$ & 0.7 & note $c$ \\[1pt] 
\hspace{0.2cm}
 $-$ Fundamental & & & & & \\ 
\hspace{0.2cm}
  Plane: $\sum_e, r_e$ & $\surd$ & $\surd$ & $\div$ & $>$0.7 & note $d$ \\[3pt] 
Via \mbh{}\,$-$\,$L_{\rm bulge}$ & & \\
\hspace{0.2cm}
  \& scaling rel.: &  & & & &  \\ 
\hspace{0.2cm}
 $-$ $M_R$  & $\surd$ & $\surd$ & $\div$ & 0.5\,$-$\,0.6 & note $e$ \\[2pt] 
\tableline \tableline
\multicolumn{6}{l}{{\sl Notes} $-$ ($a$) Improve $R$\,$-$\,$L$ relation; understand outliers, ($b$) Extend to more}\\
\multicolumn{6}{l}{ distant AGNs and luminous quasars, ($c$) Understand scatter and outliers,}\\
\multicolumn{6}{l}{($d$) Quantify and improve accuracy, ($e$) Calibrate to Reverberation masses}\\
\end{tabular}
\end{center}
\vspace{-0.5cm}
\end{table}

\section{Secondary Methods} 
Since the primary mass determination methods are either inapplicable or
impractical for more luminous and more distant AGNs and quasars, several
secondary methods have been adopted in the literature to {\sl estimate}
the central mass. These methods, summarized in Table~2, are either 
approximations to the reverberation mapping technique 
or rely on the empirical relationships between the black-hole mass, \mbh, 
and the properties of the host galaxy bulge: velocity dispersion, $\sigma_{\ast}$, 
or bulge luminosity, $L_{\rm bulge}$.  The measurements of the bulge properties 
are most useful for AGNs at low redshift ($z\lsim1$) where the host galaxy 
is easier to characterize. It is noteworthy that, even so, these methods can 
yield mass estimates that are currently quite uncertain, as explained below.

\subsection{Scaling Relationships} 
Based on single-epoch spectroscopy the scaling relationships are approximations 
to the virial-mass measurements obtained from reverberation mapping. The method 
relies on the radius -- luminosity relationship, established by reverberation 
mapping. For a photoionized BLR, its size $R$ should scale with the square root of 
the source luminosity, $R \propto L^{0.5}$. Kaspi \et (2000) found empirically 
that the size of the \hb{} emitting region scales with the continuum luminosity 
at 5100\AA{} to the power of 0.7: $R$(\hb) $\propto L_{\lambda}^{0.7}$(5100\AA). 
While this difference is yet to be understood (work is in progress; \eg 
Peterson \& Onken, this volume), this means that \mbh{} can be {\sl estimated} 
when we have measurements of the continuum luminosity and the emission-line 
width: \mbh{} $\propto$ FWHM$^2$\,$L_{\lambda}^{0.7}$.
(Note, the slope adopted in the literature is not uniform but ranges between 0.5 
and 0.7).  
The use of FWHM is a single-epoch approximation to the velocity dispersion of the 
line-emitting gas responding to continuum variations (Peterson \& Onken 2004,
this volume). 
Vestergaard (2004b, this volume) explains why these scaling relations are reasonable
\mbox{and why they can be applied to luminous, high-$z$ quasars.}

Three different broad emission lines (\hb, \mgii, and \civ) have been adopted along
with continuum luminosities (at 5100\AA{}, 3000\AA{}, and 1350\AA{}, respectively)
to estimate black-hole masses in large samples of quasars and AGNs [\hb{}, \civ{}: 
Vestergaard (2002; 2004b, this volume); \mgii: McLure \& Jarvis (2002); Jarvis \& McLure (2004, this volume)]. 
There are pros and cons to each method as outlined next.  Using optical spectroscopy, the
\hb{} method probes redshifts below 0.9, \mgii{} probes redshifts between 0.3 and
2.3, while \civ{} is accessible at redshifts between 1.2 and 4.9.  While
\feii{} emission heavily contaminates the \mgii{} line profile (\eg Francis \et 1991;
Vestergaard \& Wilkes 2001) and thus needs to be subtracted to allow measurements of
FWHM(\mgii), typically only strong \feii{} emission affects the \hb{} line width.
\civ{} is only affected in the extreme wings by \heii \lam 1640, and occasionally by
\niv \lam 1486, and (weak) \feii{} multiplets. However, the contribution to \hb{} 
from the narrow-line region needs to be subtracted. The only concern about \civ{} is the
possible presence of outflowing high-ionization BLR gas, which may yield blue asymmetric
broad profiles. This is seen very clearly in the case of narrow-line Seyfert~1 galaxies 
(\eg Leighly 2000) 
for which the \civ{} method is not well suited. But very few %, if any, 
quasars appear
to display such asymmetric profiles, so emitting outflows are probably not a general 
concern for quasars (Vestergaard 2004).
A notable assumption enters the \mgii{} method, namely that \mgii{} {\sl should} be 
emitted co-spatially with \hb{} and should therefore have similar line 
widths (which seems to be the case; McLure \& Jarvis 2002). Then
FWHM(\mgii) can be used as a direct surrogate for \hb{} in the scaling 
relation.  Unfortunately, the only published data on \mgii{} constrain the \mgii{} lag
(or the distance from which it is emitted) very poorly but the analysis by
Clavel et al.\ (1991) and Dietrich \& Kollatschny (1995) suggest that
\mgii{} is not necessarily emitted co-spatially with \hb{}.
%Unfortunately, the only published data on \mgii{} show that it is 
%emitted from a distance {\bf twice} that of \hb{} (Dietrich \& Kollatschny 1995). 
The additional advantage of the available \hb{} and \civ{} scaling relations 
is that they are calibrated to the more accurate reverberation mapping masses of
nearby AGNs (Wandel, Peterson, \& Malkan 1999; Kaspi \et 2000; Vestergaard 2002). 
The statistical scatter indicates
1$\sigma$ uncertainties of factors 2.5 to 3 (relative to the reverberation masses) 
for these scaling relations. 
Similar uncertainties are argued to be obtainable with \mgii{} (McLure \& Jarvis 2002). 
Nonetheless, individual mass estimates may be uncertain by as much as a factor 10. 
The po\-wer 
of scaling relations is in their application to large statistical AGN samples.

The uncertainties of these scaling relations are dominated by the scatter
in the $R - L$ relation, and improvements to this relation is thus desirable.  
Also, there is a need to understand the outliers in the scaling relations (Table~2). 
Dramatically improving the accuracy of the reverberation masses, on which the
scaling relations rely, will require a dedicated space-based observatory, such as 
the proposed MIDEX mission {\it Kronos}  (\eg Peterson \et 2004).

The radius $R$ can also be estimated with the reasonable assumption that the product
of the ionization parameter and the electron density is roughly similar in all
objects (the photoionization method). In this case the mass estimate is a func\-tion
of $L^{0.5}$\,FWHM$^2$ and seems accurate to within a factor two of the reverberation
masses (Wandel,\,Peterson,\,\&\,Malkan\,1999).

\subsection{Via the \mbox{\boldmath{$M_{\rm BH} - \sigma_{\ast}$}} Relationship }

As the $M_{\rm BH} - \sigma_{\ast}$ relation for quiescent galaxies is
relatively tight and AGNs also fall on this relationship, good measurements of 
the bulge velocity dispersion in AGN host galaxies can be used to estimate the
black-hole mass 
(and with a similar uncertainty; Table~2). So far, only AGNs at $z < 0.06$ have had 
$\sigma_{\ast}$ measured because the \caii \lam \lam 8498, 8542, 8662
absorption lines used for this measurement move into the atmospheric water
vapor bands at $z \geq 0.06$ (Ferrarese \et 2001).

Nelson (2000) suggested using FWHM(\oiii) as a proxy for $\sigma_{\ast}$
since the near-nuclear stellar velocity dispersion correlates with FWHM(\oiii) for a 
sample of 75 Seyfert galaxies (Nelson \& Whittle 1996).
Boroson (2003) used data from the SDSS Early Data Release 
to assess and confirm that the 1$\sigma$ uncertainty of this method is a factor of 5. 
For this method to be more useful, both the scatter and the outliers need to be 
understood (Table~2). Some line asymmetries are suspected to be connected with outflows
and can be prominent in radio sources where the narrow-line region and the radio 
source may interact.

Studies in the literature have used AGN host galaxy imaging and the Fundamental 
Plane for elliptical galaxies to measure the effective surface brightness, 
$\sum_e$, and the effective radius, $r_e$, to derive first $\sigma_{\ast}$ and 
then \mbh{}.  The reasoning is that especially nearby quasars and radio galaxies
seem to reside in giant elliptical galaxies, which {\sl should} fall on the 
Fundamental Plane. Unfortunately, this method potentially has large uncertainties
because (a) $\sum_e$ and $r_e$ are very hard to measure accurately in the presence
of the bright nucleus of most AGNs, (b) the bulge/disk decomposition process is 
difficult even for nearby quiescent galaxies and is particularly challenging for AGNs,
and (c) the method is subject to the uncertainties
of both the Fundamental Plane and the \mbh{} $- \sigma_{\ast}$ relation.
For radio galaxies, unaffected by nuclear glare, a factor 4 (or larger) uncertainty in 
estimating
$\sigma_{\ast}$ alone seems appropriate (Woo \& Urry 2002). The combined
uncertainty of the Fundamental Plane \mbox{method is thus expected to exceed a factor 5.}

\subsection{Via the \mbox{\boldmath{$M_{\rm BH} - L_{\rm bulge}$}} and Scaling Relationships} 

The black-hole mass of quiescent galaxies correlates, in addition to $\sigma_{\ast}$, 
with the bulge luminosity, $L_{\rm bulge}$, although with larger scatter. For this
reason, the \mbh{} $- L_{\rm bulge}$ relation is generally not preferred for
mass estimates of normal galaxies. This method has also proven difficult and
prone to significant uncertainties owing to the nuclear glare affecting the
bulge/disk decomposition process (\eg Wandel 2002). Nonetheless, McLure \& Dunlop (2001, 2002)
argue that with careful $R$-band measurements of the AGN host galaxy, first
$L_{\rm bulge}$ and then \mbh{} can be estimated to within a factor 3 to 4.
Unfortunately, this method has not yet been calibrated to reverberation mapping
masses, but rather to mass estimates based on scaling laws. This may not be a
severe problem since the authors find their AGNs to have the same slope and scatter
as the inactive galaxies with dynamical mass measurements to within the errors.

\section{Future Efforts} 
%\vspace{-0.15cm}
Table~2 lists suggestions to how we may improve the 
mass estimation methods. 
In addition, it is important to understand their limitations as 
well as their efficacy. 
Since reverberation masses provide the anchor of the scaling relations a few
comments thereon are in order. An understanding of the BLR structure and
kinematics is required to determine the absolute zero-point of the 
reverberation masses (the `$f$' factor). This is a non-trivial but an achievable
task with a dedicated observatory like {\it Kronos} and existing advanced analysis 
techniques (Peterson \et 2004; Peterson \& Horne 2004). 
Also, direct comparisons of stellar dynamics and reverberation 
mapping need to be effected.  Moreover, the odd objects need to be understood:\
for example, for 3C\,390.3 the lower ionization lines appear to have shorter 
lags, in contrast to other AGNs. As this object is the only broad-line 
radio source which has been monitored, it needs to be established whether or 
not it belongs to a separate class of objects. Understanding this behavior is 
also \mbox{important for our understanding of the typical BLR structure.}

%\vspace{-0.4cm}

\end{document}